\documentclass{article}
\usepackage{spconf,amsmath,graphicx}
\usepackage[font=small,skip=0pt]{caption}
\usepackage{amsfonts}
\usepackage{wrapfig}

\title{learning discriminative features from spectrograms using center loss for speech emotion recognition}

 \name{Dongyang Dai$^{1, 2}$, Zhiyong Wu$^{1,2,3,*}$\thanks{* Corresponding author}, Runnan Li$^{1,2}$, Xixin Wu$^3$, Jia Jia$^{1,2}$, Helen Meng$^{1,3}$}
 
 \address{ $^1$Tsinghua-CUHK Joint Research Center for Media Sciences, Technologies and Systems, \\
 	 Graduate School at Shenzhen, Tsinghua University, Shenzhen, China\\
 	 $^2$Tsinghua National Laboratory for Information Science and Technology (TNList),\\
 	 Department of Computer Science and Technology, Tsinghua University, Beijing, China \\
 	 $^3$Department of Systems Engineering and Engineering Management, \\
 	 The Chinese University of Hong Kong, Shatin, N.T., Hong Kong SAR, China\\
 	 \{ddy17,lirn15\}@mails.tsinghua.edu.cn, \{zywu,wuxx,hmmeng\}@se.cuhk.edu.hk, jjia@tsinghua.edu.cn}
\begin{document}

%
\maketitle
\begin{abstract}
Identifying the emotional state from speech is essential for the natural interaction of the machine with the speaker. However, extracting effective features for emotion recognition is difficult, as emotions are ambiguous. We propose a novel approach to learn discriminative features from variable length spectrograms for emotion recognition by cooperating softmax cross-entropy loss and center loss together. The softmax cross-entropy loss enables features from different emotion categories separable, and center loss efficiently pulls the features belonging to the same emotion category to their center. By combining the two losses together, the discriminative power will be highly enhanced, which leads to network learning more effective features for emotion recognition. As demonstrated by the experimental results, after introducing center loss, both the unweighted accuracy and weighted accuracy are improved by over 3\% on Mel-spectrogram input, and more than 4\% on Short Time Fourier Transform spectrogram input.
\end{abstract}
\section{introduction}

Speech emotion recognition (SER) is crucial for natural human-computer interaction. An SER system extracts features from the speech waveform and then classifies them into the corresponding emotion categories. And how to extract features containing enough emotional information has drawn growing interest.

For SER, traditional methods extract frame-level features from overlapped frames on speech signals and apply statistic functions on them to get additional features \cite{el2011survey}. Since deep neural network (DNN) can learn high-level invariant features from raw data \cite{bengio2013representation} and deep learning brings a lot of breakthroughs in many fields \cite{lecun2015deep}, more and more methods utilizing neural networks to extract valid features from raw data have emerged. In \cite{han2014speech}, DNN and extreme learning machine were utilized to extract high-level features from low-level features. A bi-directional Long Short-Term Memory model was used in \cite{lee2015high}  to extract high level feature representations for SER. In \cite{trigeorgis2016adieu}, representation learning was performed on raw waveform for end-to-end SER. Convolutional and recurrent neural networks were applied to learn high-level representations from spectrograms in SER task \cite{satt2017efficient}.

Emotions are naturally ambiguous \cite{mower2009interpreting}, different types of emotions might be confusing, increasing the difficulty of extracting effective features \cite{chao2016long}. A trending methodology to release the ambiguity of emotion is to design an appropriate loss function instructing the neural network to learn discriminative features which have smaller intra-class variance and larger inter-class variance. A ``pairwise discriminative task'' was introduced in \cite{lian2018pairwise} to learn the similarity and distinction between two audios, which utilized cosine similarity loss together with binary cross entropy loss. In the task, pairwise audios were fed into an audio encode networks to extract audio vectors, and a following discriminative network judged whether the pairwise audios belong to the same emotion category using binary cross-entropy loss. The extracted audio vectors from the same class were made ``close'' by the effect of cosine similarity loss and they were classified by a support vector machine ({SVM}). In \cite{huang2018speech},  a triplet framework was proposed to extract discriminative features by using triplet loss\cite{schroff2015facenet}, whose input was triplets including two utterances from the same emotion class and one utterance from other classes. Then, similarly as \cite{lian2018pairwise}, an SVM fed with extracted features was used for classifying.

Recent methods to learn discriminative features for SER via using cosine similarity loss\cite{lian2018pairwise} or triplet loss\cite{huang2018speech} adopt a two-step strategy. These methods extract discriminative features in one step, and classify features with SVM in the other step. However, the two-step strategy may bring a reduction of SER performance as targets of the two steps may not be completely consistent. Besides, the performance of these methods depends heavily on the selection strategy of pairwise audios or triplets. In this paper, we propose a novel approach to extract discriminative features for SER from variable length spectrograms in an end-to-end manner using a jointly supervised loss consisting of softmax cross-entropy loss and center loss\cite{wen2016discriminative}. Center loss pulls features in the same class closer to their class center, and softmax cross-entropy loss separates features from different emotional categories. Through optimizing center loss together with softmax cross-entropy loss, discriminative features will be learned for better SER results. Compared with cosine similarity loss method \cite{lian2018pairwise} and triplet loss method \cite{huang2018speech}, center loss could naturally be integrated in common SER models, which dispense constituting sample pairs or sample triplets and the additional use of SVM classifier.

\section{the proposed approach}



\begin{figure}[htb]
	
	\begin{minipage}[b]{1.0\linewidth}
		\centerline{\includegraphics[width=8.5cm]{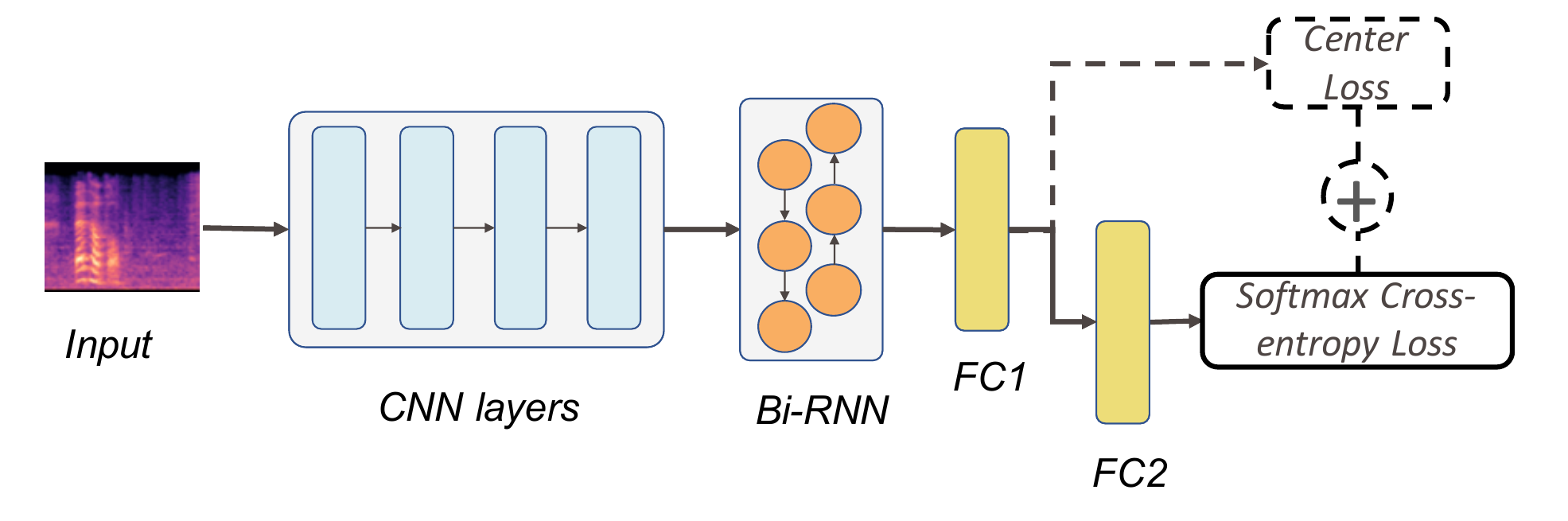}}
	\end{minipage}
	\caption{Model framework. The model takes variable length spectrograms as input and learns discriminative features for SER.}
	\label{fig:model-framework}
\end{figure}

Fig.\ref{fig:model-framework} depicts the framework of the proposed model, which includes several 2-D Convolutional Neural Network layers ({CNN layers}), a bidirectional Recurrent Neural Network layer ({Bi-RNN}) and two fully-connected layers (FC1 and FC2). Softmax cross-entropy loss and center loss are utilized in our model.

CNN layers extract spatial information from a variable length spectrogram to get a variable length sequence, Bi-RNN compresses the variable length sequence down to a fixed-length vector. FC1 projects Bi-RNN's output to the desired dimensionality. FC2, whose output denotes the posterior class probabilities, is used to calculate softmax cross-entropy loss. Softmax cross-entropy loss enables the network to learn separable features, and center loss reduces features' intra-class variation simultaneously.
\subsection{Model details}
\label{ssec:model-detail}
The model takes a Short Time Fourier Transform ({STFT}) spectrogram or Mel-spectrogram as input, whose size is $L_T \times L_F$. $L_T$ is variable depending on the length of audio, and $L_F$ is the dimension related to the frequency domain.

\begin{figure}[htb]
	\begin{minipage}[b]{0.04\linewidth}
		\hfill
	\end{minipage}
	\begin{minipage}[b]{0.52\linewidth}
		\centering
		\centerline{\includegraphics[width=5.0cm]{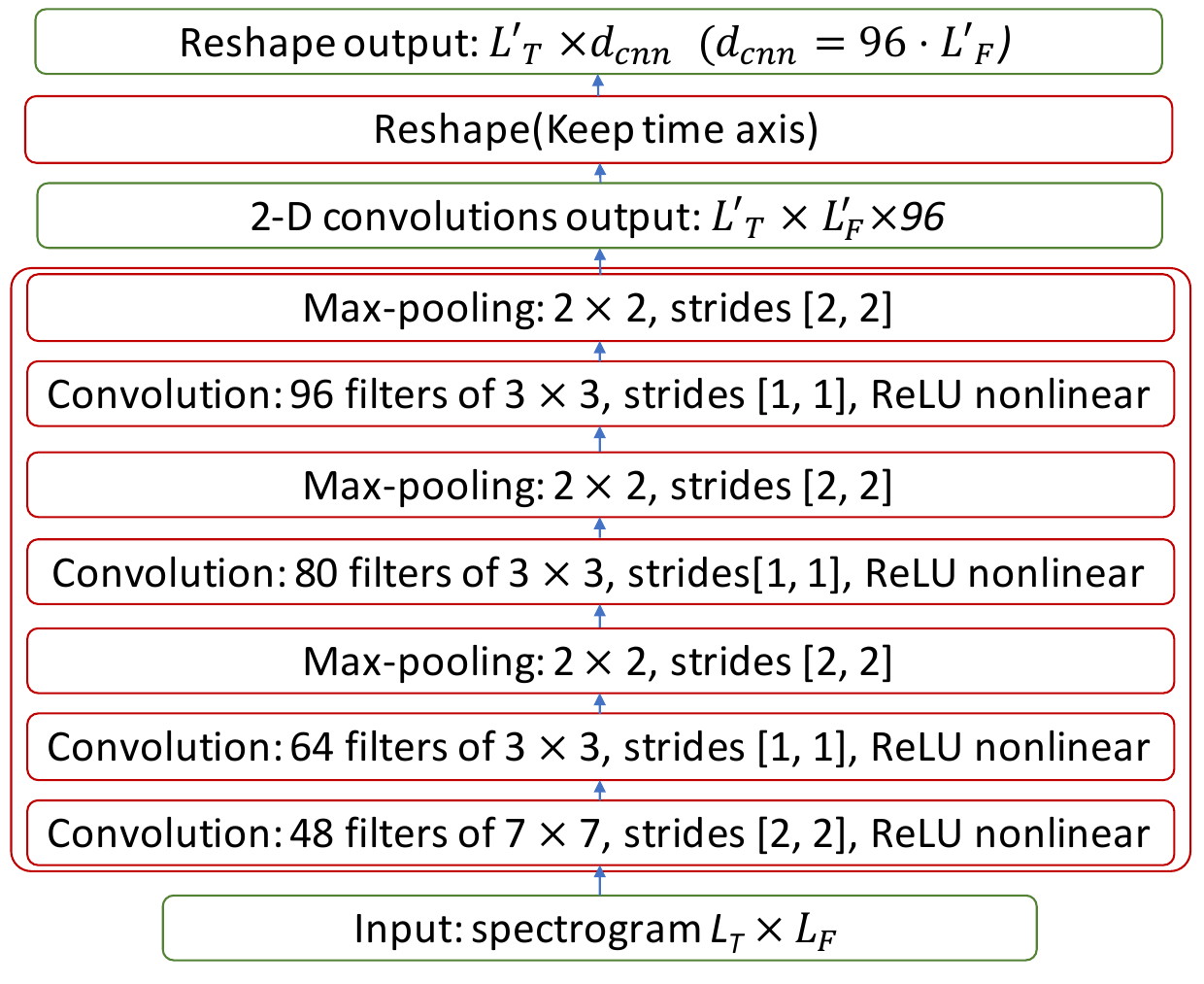}}
		\centerline{\small(a)}\medskip
		
	\end{minipage}
	\hfill
	\begin{minipage}[b]{.42\linewidth}
		\centering
		\centerline{\includegraphics[width=4.2cm, angle=270]{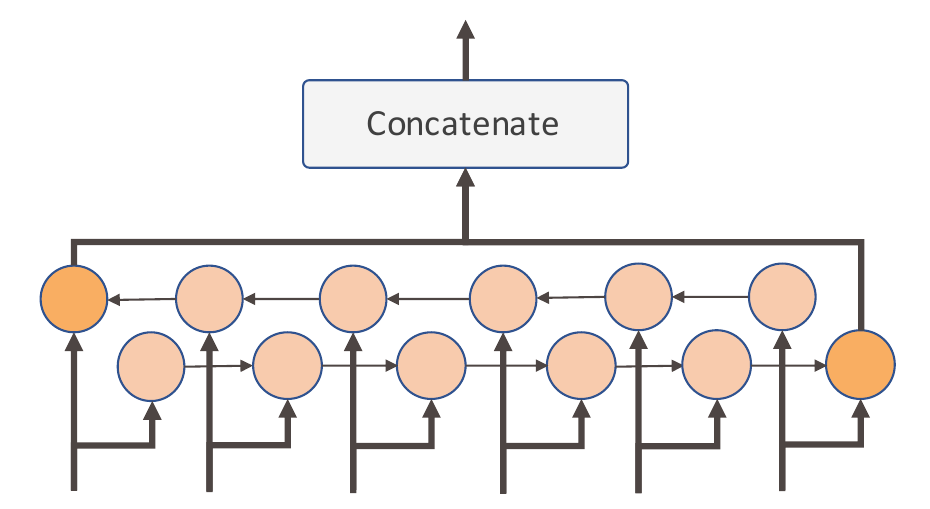}}
		\centerline{\small(b)  }\medskip
	\end{minipage}
	\caption{Model Details.(a) Model details of CNN layers, (b) Bi-RNN compresses variable length sequence to a fixed-length vector}
	\label{fig:model_detail}
\end{figure}

According to the experience of computer vision, the convolutional network, whose first layer uses large convolution kernels and the remaining layers use small convolution kernels, perform well \cite{simonyan2014very, He2015Deep}. Besides, after dozens of tests, we determined the details of CNN layers as Fig.\ref{fig:model_detail}-(a).

The Bi-RNN compresses variable length sequence produced by CNN layers to a fixed-length vector by concatenating the last output of forward RNN and backward RNN, as shown in Fig.\ref{fig:model_detail}-(b). Bi-RNN is implemented with 128-width Gated Recurrent Unit ({GRU})\cite{Cho2014Learning}, so the dimension of Bi-RNN's output is 256.

FC1 projects Bi-RNN's output to the desired feature space of target dimension $d$ ($d=64$ in our experiments), with PReLU\cite{He2015Delving} activation function. We take FC1's output $z \in \mathbb{R}^d$ as the learned feature and calculate center loss according to $z$. The output of FC2 denotes the predicted posterior probabilities to corresponding emotion categories. And the parameters in FC2 is used for calculating softmax cross-entropy loss, which will be described in detail in section~\ref{ssec:cross-entropy}. 

\subsection{Softmax cross-entropy loss}
\label{ssec:cross-entropy}

Softmax cross-entropy loss instructs the model to learn separable features, and it is common in multi-classification tasks. The softmax loss function is presented as equation~\ref{eq:cross-entropy}:
\begin{equation} \label{eq:cross-entropy}
L_s^0 = - \frac{1}{m}\sum_{i=1}^{m}{log(\frac{e^{W_{y_i}^\mathrm{T}z_i+b_{y_i}}}{\sum_{j=1}^{n}e^{W_j^\mathrm{T}z_i+b_j}})}
\end{equation}
where $m$ means the size of mini-batch and $n$ is the number of emotion categories. $z_i \in \mathbb{R}^d$ is the $i$-th deep feature, belonging to $y_i$-th emotion category ($y_i \in  \{1, 2, ... , n\}$). $W_j \in \mathbb{R}^d$ denotes the $j$-th column of the weights $ W \in \mathbb{R}^{d \times n} $ in FC2, $b \in \mathbb{R} ^ n $ is the bias in FC2 and $ b_j $ is the $j$-th term of $b$.

\subsection{Center loss}
To reduce intra-class variation of learned features, we introduce center loss in our model. Our model keeps a global center for each class and pulls features closer to their corresponding centers. The formula of center loss is given as follows:
\begin{equation} \label{eq:center-loss}
L_c^0=\frac{1}{m}\sum_{i=1}^{m}||z_i - c_{y_i}||^2
\end{equation}
where $c_j$ ($j \in \{1, 2, ... , n\}$) denotes the global class center of features corresponding to the $j$-th emotion category. Through optimizing the center loss, the distance between features within the same class becomes smaller. $c_j$ is initialized with 0 and updated per mini-batch iteration based on $\dot{c}_j$, which is the $j$-th class center of features from a mini-batch, caculated by equation~\ref{eq:batch-center} when $\sum_{i=1}^{m}\delta(y_i = j) > 0$. $\delta(condition) = 1$ if the $condition$ is satisfied, otherwise $\delta(condition) = 0$.
\begin{equation} \label{eq:batch-center}
\dot{c}_j = \frac{\sum_{i=1}^{m}\delta(y_i = j) z_i}{\sum_{i=1}^{m}\delta(y_i = j)}
\end{equation}

The global class center $c_j$ is updated as equation~\ref{eq:center-update}. $\alpha$ is a hyperparameter controlling the update rate of $c_j$ when there are features corresponding to $j$-th emotion category in the new mini-batch, while $c_j$ keeps its previous value when no corresponding features in the new mini-batch. $c^t_j$ and $\dot{c}^t_j$ denotes the $t$-th iteration's value of  $c_j$ and $\dot{c}_j$ respectively.
\begin{equation} \label{eq:center-update}
c_j^{t+1}=\left\{
\begin{array}{lr}
(1 - \alpha) c_j^t + \alpha \dot{c}^t_j & \sum_{i=1}^{m}\delta(y_i = j) > 0 \\
c_j^t & \sum_{i=1}^{m}\delta(y_i = j) = 0
\end{array}
\right.
\end{equation}
\subsection{Weighted loss and joint loss}

Because of class imbalance, instead of using $L_s^0$ and $L_c^0$ directly, we assigned weights to softmax cross-entropy loss and center loss in our experiments, shown in equation~\ref{eq:weighted-cross-entropy} and equation~\ref{eq:weighted-center-loss}. The weight $\omega_j$($j \in \{1, 2, ... , n\}$)  is  in inverse proportion to the sample number of the $j$-th class in training set.
\begin{equation} \label{eq:weighted-cross-entropy}
L_s = - \frac{1}{\sum_{i=1}^{m}\omega_{y_i}}\sum_{i=1}^{m}{\omega_{y_i}log(\frac{e^{W_{y_i}^\mathrm{T}z_i+b_{y_i}}}{\sum_{j=1}^{n}e^{W_j^\mathrm{T}z_i+b_j}})}
\end{equation}
\begin{equation} \label{eq:weighted-center-loss}
L_c=\frac{1}{\sum_{i=1}^{m}\omega_{y_i}}\sum_{i=1}^{m}{\omega_{y_i}||z_i - c_{y_i}||^2}
\end{equation}

Our neural network is trained using a joint loss comprised of softmax cross-entropy loss and center loss, calculated as equation~\ref{eq:joint-loss}. $\lambda$ is a hyperparameter trading off center loss against softmax cross-entropy loss. When  $\lambda=0$, the model is trained using only softmax cross-entropy loss.
\begin{equation} \label{eq:joint-loss}
L=L_s+ \lambda L_c
\end{equation}
%
%
%

\section{experiments and analysis}

\subsection{Experimental setup}
Our model was tested on the Interactive Emotional Dyadic Motion Capture (IEMOCAP) \cite{busso2008iemocap} dataset, which was designed for studying multimodal expressive dyadic interactions. It contains approximately 12 hours of audiovisual data, including video, speech, motion capture of face and text transcriptions. For training and evaluation, we used categorical emotions neutral, angry, happy, sad and excited which represent the majority of the emotion categories in the database (5531 utterances), happy and excited were merged since they are close in the activation and valence domain \cite{busso2008iemocap}.

As there is data imbalance between classes -- neutral (30.9\% of the total dataset), angry (19.9\%), happy (29.6\%), and sad (19.6\%), we adopted both the unweighted accuracy ({UA}, the mean value of the recall for each class) and the weighed accuracy ({WA},  the number of correctly classified samples divided by the total amount of samples) as metrics to evaluate SER performance.  In our experiments, the dataset was divided into 5 subsets randomly keeping the emotion distribution, 4 subsets were used for training, half of the last subset was used as development set and half as test set. We repeated cross-validation  5 times to get the final average result.

Our experiments were conducted on log scale Mel-spectrogram and log scale STFT spectrogram respectively. To get spectrogram, a sequence of overlapping Hamming windows were applied to the speech signal, with window shift of 10msec, and window size of 40msec. The speech signal was sampled at 16kHz and the DFT length was 1024. The number of Mel bands was 128 when calculating Mel-spectrogram. We assumed that 14s long utterance contains enough emotional information. So for utterance whose duration is longer than 14s (2.07\% of the total dataset), we only extracted the middle 14s to calculate spectrogram. 

During the training phase, we used Adam \cite{kingma2014adam} optimizer, set learning rate to 0.0003, and the size of mini-batch was 32. We applied the parameters maximizing the UA of the development set as the model's final parameters.

\subsection{Experiments on Mel-Spectrogram}

\begin{figure}[htb]	
	\begin{minipage}[b]{0.48\linewidth}
		\label{fig:metric_lambda_alpha_a}
		\centering
		\centerline{\includegraphics[width=4cm, height=2.8cm]{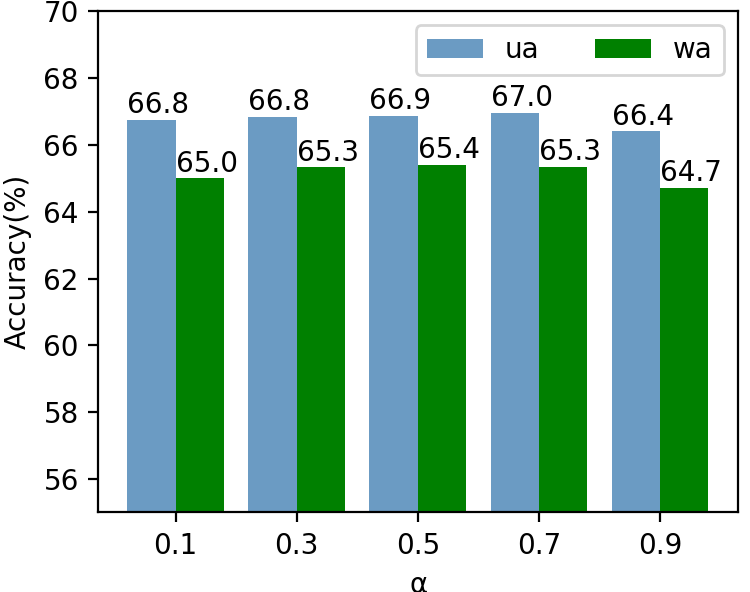}}
		\centerline{\small(a)}\medskip
		
	\end{minipage}
	\hfill
	\begin{minipage}[b]{.48\linewidth}
		\centering
		\centerline{\includegraphics[width=4cm,height=2.8cm]{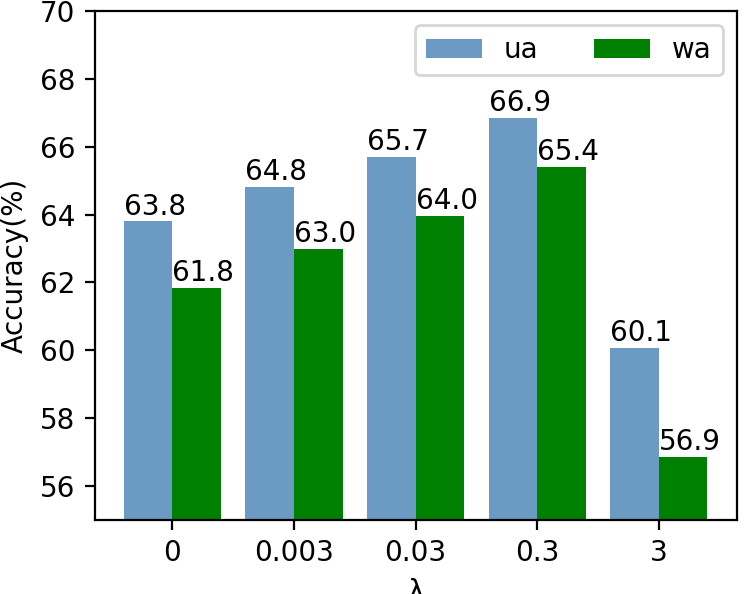}}
		\centerline{\small(b)  }\medskip
		\label{fig:metric_lambda_alpha_b}
	\end{minipage}
	\caption{UA and WA on log scale Mel-spectrogram input.(a) model with different $\alpha$ when fixing $\lambda$ = 0.3 (b) model with different $\lambda$ when fixing $\alpha$ = 0.5}
	\label{fig:metric_lambda_alpha}
	\vspace{-10pt}
\end{figure}

As $\alpha$ controls the update rate of class centers and $\lambda$ dominates the weight of center loss, we conducted experiments to investigate the effect of hyperparameter $\alpha$ and $\lambda$ on Mel-spectrogram input. The  experimental results are shown in Fig.\ref{fig:metric_lambda_alpha}. Fig.\ref{fig:metric_lambda_alpha}-(a) illustrates that the UA and WA  are not sensitive to $\alpha$, Fig.\ref{fig:metric_lambda_alpha}-(b) demonstrates that the SER performance can be significantly improved with proper value of $\lambda$. When $\lambda=0$ ({setting 1}), the UA is 63.80\% and the WA is 61.83\%. The UA and WA is 66.86\% and 65.40\% respectively when $\lambda=0.3, \alpha=0.5$ ({setting 2}). The UA and WA are both increased by over 3\% when using center loss with proper hyperparameters.

\begin{figure}[htb]
	\begin{minipage}[b]{.48\linewidth}
		\centering
		\centerline{\includegraphics[width=4.2cm]{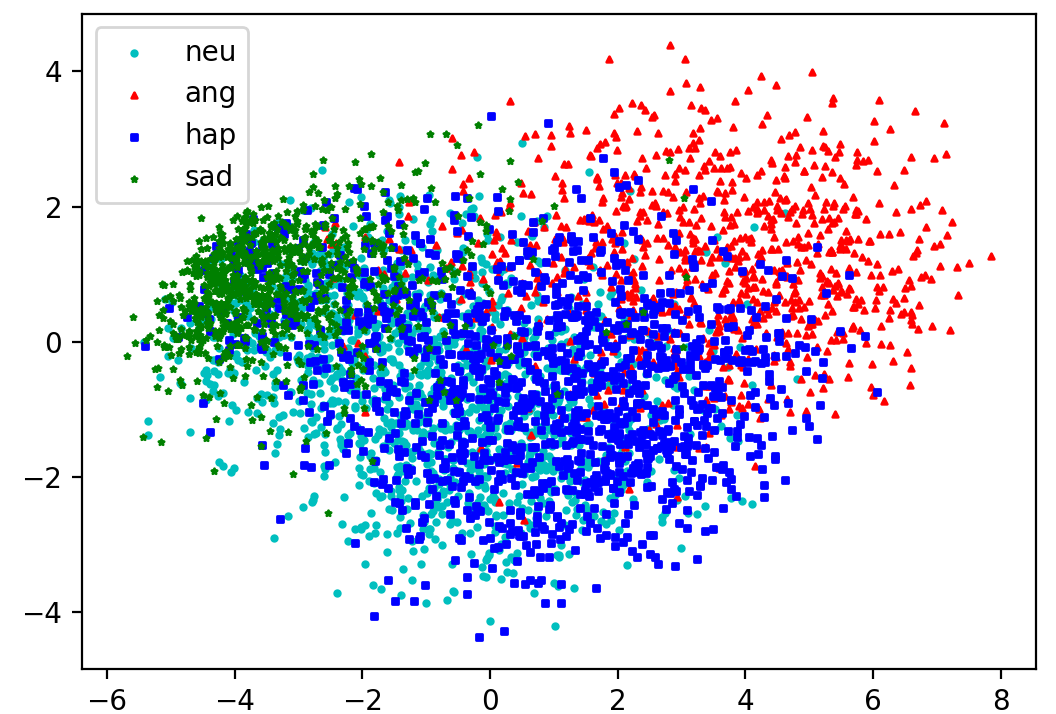}}
		\centerline{\small(a)  }\medskip
	\end{minipage}
	\hfill
	\begin{minipage}[b]{0.48\linewidth}
		\centering
		\centerline{\includegraphics[width=4.2cm]{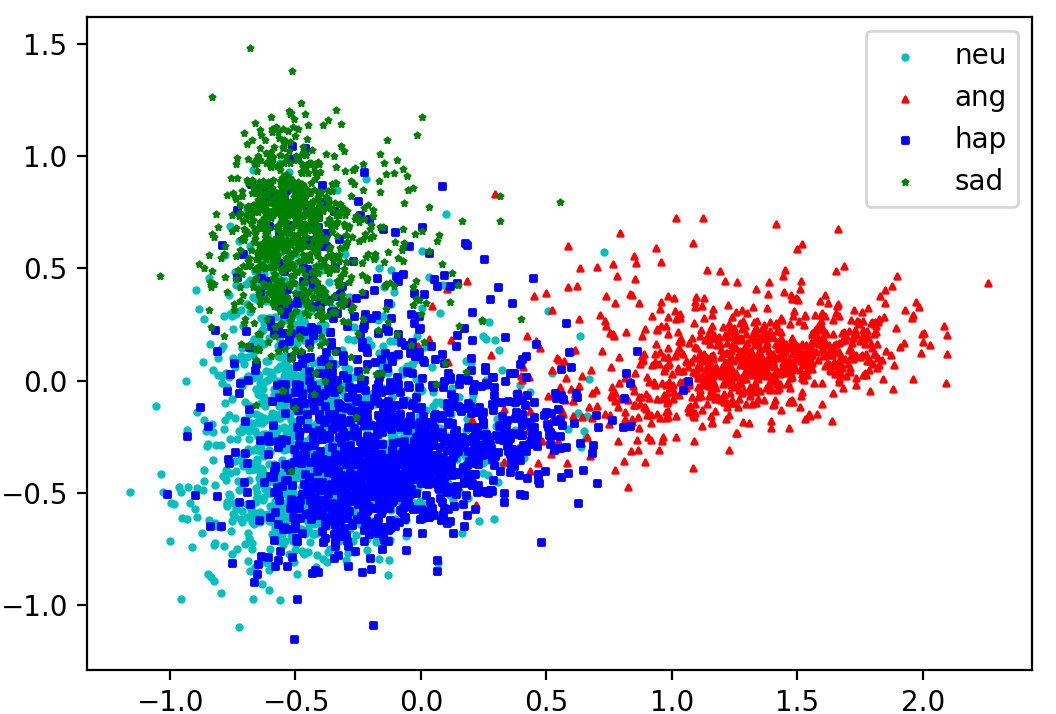}}
		\centerline{\small(b)}\medskip
	\end{minipage}
	
	\begin{minipage}[b]{.48\linewidth}
		\centering
		\centerline{\includegraphics[width=4.2cm]{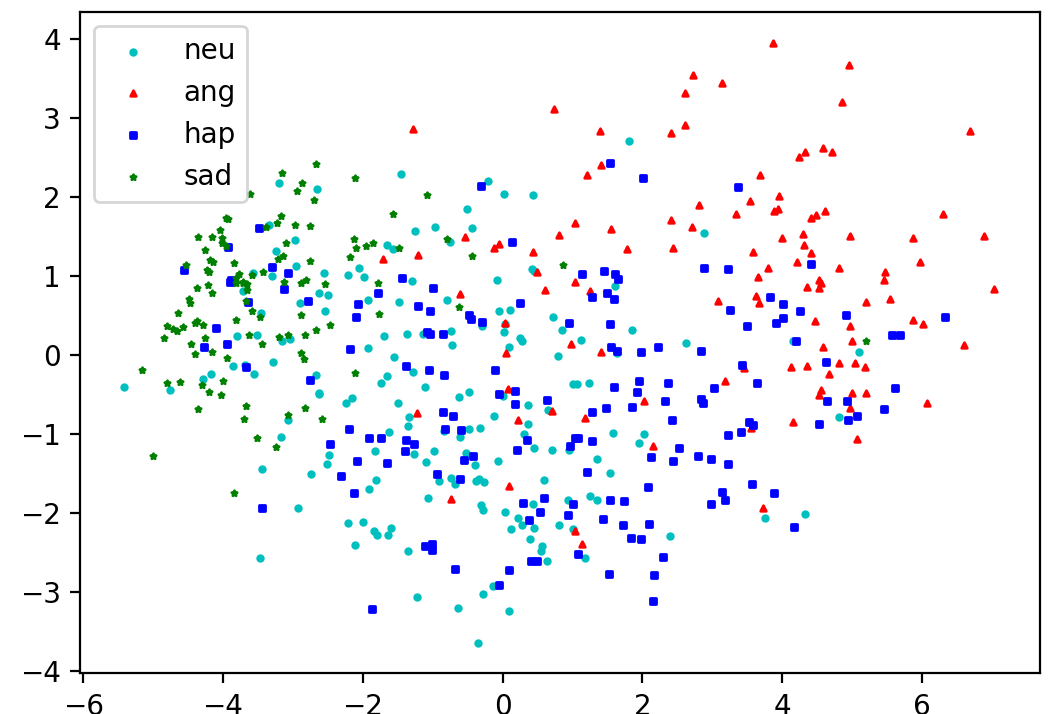}}
		\centerline{\small(c)  }\medskip
	\end{minipage}
	\hfill
	\begin{minipage}[b]{0.48\linewidth}
		\centering
		\centerline{\includegraphics[width=4.2cm]{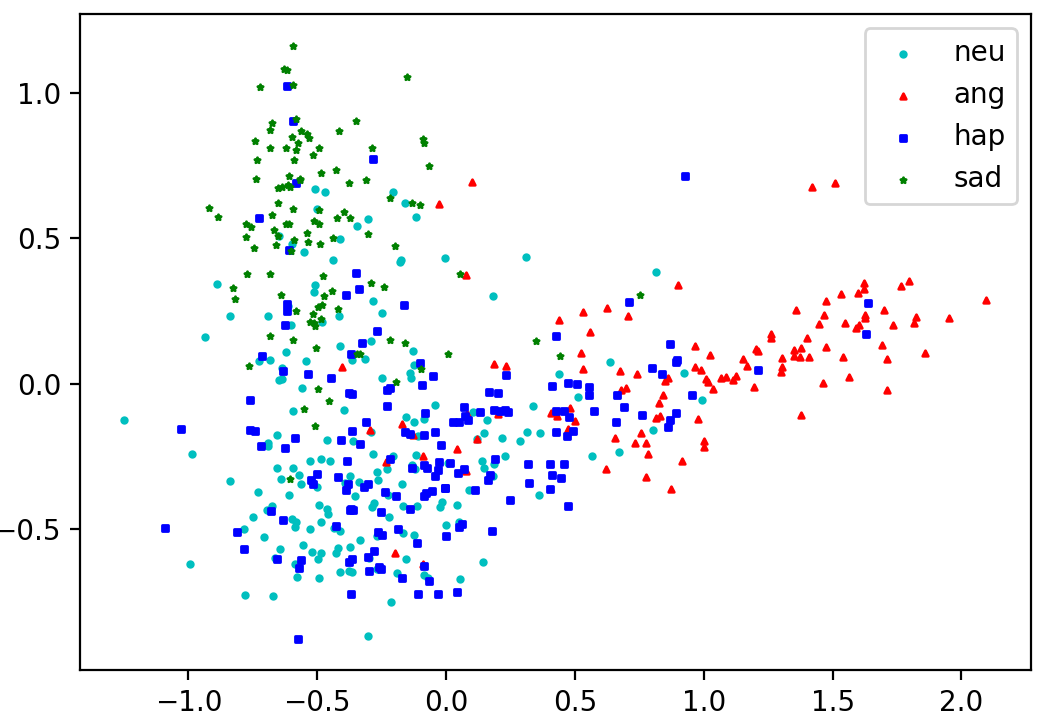}}
		\centerline{\small(d)}\medskip
	\end{minipage}
	\caption{PCA embedding of features from, (a) training set on setting 1, (b) training set on setting 2, (c) test set on setting 1, (d) test set on setting 2}
	\label{fig:pca}
	\vspace{-10pt}
\end{figure}

\begin{figure}[htb]
	\begin{minipage}[c]{0.55\linewidth}
		\centerline{\includegraphics[width=4.5cm]{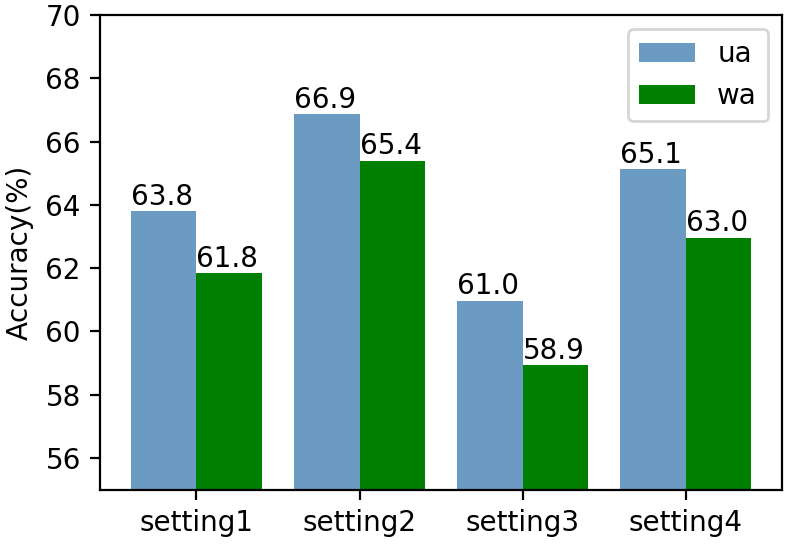}}
	\end{minipage}
	\begin{minipage}[c]{0.36\linewidth}
			\scalebox{0.65}{
			\begin{tabular}{c|cc}
						\hline
						& Hyperparameters&Input\\
			
						\hline
						setting1 & $\lambda$=0 & Mel  \\
						setting2 & $\lambda$=0.3, $\alpha$=0.5 & Mel  \\
						setting3 & $\lambda$=0 & STFT\\
						setting4  & $\lambda$=0.3, $\alpha$=0.5 & STFT \\
						\hline
			\end{tabular}}
	\end{minipage}
	\caption{The UA and WA on setting 1 $\sim$ setting 4.}
	\label{fig:setting1-4}

\end{figure}

To illustrate the discriminative power provided by center loss, we applied Principal Component Analysis({PCA}) to embed learned features. The PCA embeddings (of features produced by once experiment in cross-validations on setting 1 and on setting 2 respectively) are drawn in Fig.\ref{fig:pca}. Comparing Fig.\ref{fig:pca}-(b) with Fig.\ref{fig:pca}-(a) or Fig.\ref{fig:pca}-(d) with Fig.\ref{fig:pca}-(c), we could find that features belonging to the same class are more compact when using center loss. After introducing center loss together with softmax cross-entropy loss to train the model, the discriminative power can be enhanced, which leads to the model learning more effective features for SER.

The final confusion matrix was calculated by averaging confusion matrices from 5 times cross-validations, the results on setting 1 and setting 2 are shown in table~\ref{table:result_log_mel}-(a) and table~\ref{table:result_log_mel}-(b). As can be seen, after introducing center loss for enhancing discriminative power, the accuracy of each emotion has been improved to different degrees.

\begin{table} [!htp]
		\caption{confusion matrix on, (a) setting 1, (b) setting 2}
		\label{table:result_log_mel}
\begin{minipage}[b]{.48\linewidth}
\centerline{
	\scalebox{0.70}{
		\begin{tabular}{c|cccc}
			\hline
			& neu & ang & hap & sad \\
			\hline
			neu & 0.575 &  0.095 & 0.164 & 0.166 \\
			ang & 0.119 & 0.691 & 0.155 & 0.035\\
			hap & 0.211 & 0.162 & 0.511 & 0.115\\
			sad  & 0.138 & 0.026 & 0.060 & 0.776\\
			\hline
		\end{tabular}}}
		\centerline{\small(a)}\medskip
\end{minipage}
		\hfill
\begin{minipage}[b]{0.48\linewidth}
\centerline{
	\scalebox{0.70}{
		\begin{tabular}{c|cccc}
			\hline
			& neu & ang & hap & sad \\
			\hline
			neu& 0.637 &  0.067 & 0.167 & 0.127 \\
			ang & 0.108 & 0.705 & 0.167 & 0.020\\
			hap & 0.219 & 0.131 & 0.556 & 0.094\\
			sad  & 0.128 & 0.025 & 0.070 & 0.777\\
			\hline
		\end{tabular}}}
		\centerline{\small(b)}\medskip
\end{minipage}
\vspace{-10pt}
\end{table}
	
\begin{table} [!htp]
	\caption{confusion matrix on, (a) setting 3, (b) setting 4}
	\label{table:result_stft}
	\begin{minipage}[b]{.48\linewidth}
		
		\centerline{
			\scalebox{0.70}{
				\begin{tabular}{c|cccc}
					\hline
					& neu & ang & hap & sad \\
					\hline
					neu & 0.544 & 0.093 & 0.185 & 0.177 \\
					ang & 0.127 & 0.681 & 0.167 & 0.025\\
					hap & 0.216 & 0.186 & 0.476 & 0.122\\
					sad  & 0.161 & 0.039 & 0.062 & 0.737\\
					\hline
				\end{tabular}}}
				\centerline{\small(a) }\medskip
	\end{minipage}
			\hfill
	\begin{minipage}[b]{0.48\linewidth}
				\centerline{
					\scalebox{0.70}{
						\begin{tabular}{c|cccc}
							\hline
							& neu & ang & hap & sad \\
							\hline
							neu & 0.573 & 0.073 & 0.196 & 0.157 \\
							ang & 0.103 & 0.720 & 0.153 & 0.022\\
							hap & 0.205 & 0.161 & 0.518 & 0.114\\
							sad  & 0.125 & 0.028 & 0.053 & 0.793\\
							\hline
						\end{tabular}}}
						\centerline{\small(b)}\medskip
	\end{minipage}
	\vspace{-10pt}
\end{table}				
				
\subsection{Experiments on STFT spectrogram}
We conducted experiments to prove that introducing center loss is also useful for learning effective features for SER on STFT spectrograms input, by comparing experiment result when $\lambda=0$ ({setting 3}) with $\lambda=0.3, \alpha=0.5$ ({setting 4}). The final confusion matrices are shown in table~\ref{table:result_stft}-(a) and table~\ref{table:result_stft}-(b), which show that each emotion's accuracy has been improved after introducing center loss. The UA and WA are 60.98\% and 58.93\% on setting 3, while 65.13\% and 62.96\% on setting 4. Both the UA and WA are improved by more than 4\% after introducing center loss.

We presented the UA and WA on setting 1 $\sim$ setting 4 shown in Fig.\ref{fig:setting1-4}. We can conclude that introducing center loss could effectively improve the SER performance by comparing setting 2 with setting 1 or setting 4 with setting 3 in Fig.\ref{fig:setting1-4}.  By comparing experimental results on setting 1 with setting 3 or setting 2 with setting 4 in Fig.\ref{fig:setting1-4}, we could find that learning effective features for SER on Mel-spectrogram input, which reduces the dimension based on human hearing characteristics, is easier than on STFT spectrogram input.

\section{conclusions}
In this paper, we presented an approach to learn discriminative features from variable length spectrograms by integrating center loss in the SER model. The 2-D PCA embedding illustrated the discriminative power when using center loss, which enables the neural network to learn more effective features for SER. Our experiment results demonstrated that center loss with proper hyperparameters is useful for improving the performance of SER on both Mel-spectrogram input and STFT spectrogram input. As center loss mainly focuses on reducing the intra-class variation of features, in future work, we will explore more in the loss function, hoping to increase the features' inter-class variation to further improve the SER performance. 

\vfill

\noindent \small \textbf{Acknowledgements}: This work is supported by National Natural Science Foundation of China (NSFC) (61433018, 61375027), joint research fund of NSFC-RGC (Research Grant Council of Hong Kong) (61531166002, N CUHK404/15) and National Social Science Foundation of China (13\&ZD189).
\pagebreak

\bibliographystyle{IEEEbib}
\bibliography{strings3,refs3}

\end{document}